\documentclass[12pt]{article}
\catcode`\@=11
\@addtoreset{equation}{section}

\global\arraycolsep=2pt
\usepackage{mathrsfs,amsbsy,amssymb,latexsym,amsfonts,amsmath,cite}
\usepackage{graphicx,color}
\usepackage{physics}
\usepackage{mathtools}
\usepackage{ulem}

\usepackage{mathrsfs,amsbsy,amssymb,latexsym,amsfonts,amsmath,cite,bm}
\usepackage{graphicx,color}
\usepackage{mathtools}
\usepackage{enumerate}

\DeclareFontFamily{U}{BOONDOX-calo}{\skewchar\font=45 }
\DeclareFontShape{U}{BOONDOX-calo}{m}{n}{
  <-> s*[1.05] BOONDOX-r-calo}{}
\DeclareFontShape{U}{BOONDOX-calo}{b}{n}{
  <-> s*[1.05] BOONDOX-b-calo}{}
\DeclareMathAlphabet{\mathcalboondox}{U}{BOONDOX-calo}{m}{n}
\SetMathAlphabet{\mathcalboondox}{bold}{U}{BOONDOX-calo}{b}{n}
\DeclareMathAlphabet{\mathbcalboondox}{U}{BOONDOX-calo}{b}{n}

\newcommand{\no}{\nonumber}

\newcommand{\cA}{\mathcal A}

\newcommand{\cG}{\mathcal G}

\newcommand{\cL}{\mathcal L}
\newcommand{\cM}{\mathcal M}
\newcommand{\cO}{\mathcal O}

\newcommand{\ch}{\mathcalboondox h}
\newcommand{\cg}{\mathcalboondox g}

\newcommand{\pa}{\partial}

\allowdisplaybreaks

\begin{document}

\renewcommand{\thefootnote}{\fnsymbol{footnote}}

\begin{flushright}
KUNS-2910
\end{flushright}
\vspace*{0.5cm}

\begin{center}
{\Large \bf  Non-Abelian Toda field theories\\[5pt]
\,from a 4D Chern-Simons theory}
\vspace*{2cm} \\
{\large  Osamu Fukushima$^{\sharp}$\footnote{E-mail:~osamu.f@gauge.scphys.kyoto-u.ac.jp},
Jun-ichi Sakamoto$^{\dagger}$\footnote{E-mail:~sakamoto@ntu.edu.tw},
and Kentaroh Yoshida$^{\sharp}$\footnote{E-mail:~kyoshida@gauge.scphys.kyoto-u.ac.jp}} 
\end{center}

\vspace*{0.4cm}

\begin{center}
$^{\sharp}${\it Department of Physics, Kyoto University, Kyoto 606-8502, Japan.}
\end{center}
\begin{center}
$^{\dagger}${\it Department of Physics and Center for Theoretical Sciences, National Taiwan University, Taipei 10617, Taiwan}
\end{center}

\vspace{2cm}

\begin{abstract}
We derive non-abelian Toda field theories (NATFTs) from a 4d Chern-Simons (CS) theory with two order defects 
by employing a certain asymptotic boundary condition. 
The 4d CS theory is characterized by a meromorphic 1-form $\omega$\,.
We adopt $\omega$ with two simple poles and no zeros, and 
each of the order defects is located at each pole. 
As a result, an anisotropy parameter $\beta^2$ can be identified with the distance between the two defects.
As examples, we can derive the (complex) sine-Gordon model and the Liouville theory.
\end{abstract}

\setcounter{footnote}{0}
\setcounter{page}{0}
\thispagestyle{empty}

\newpage

\tableofcontents

\renewcommand\thefootnote{\arabic{footnote}}

\section{Introduction}
Integrable field theories have been investigated for decades, while it is generally difficult to determine whether a given theory is integrable or not.
To prove the integrability, one needs to find a Lax pair $\cL$ which satisfies the on-shell flatness condition.
Recently, Costello, Witten and Yamazaki proposed  how to describe a lot of  integrable systems in a unified way based on a 4d gauge theory, referred to as a 4d Chern-Simons (CS) theory \cite{Costello:2013zra,CWY1,CWY2,CY}.
The 4d CS theory includes a meromorphic 1-form $\omega$\,, and appropriate boundary conditions are imposed at the poles of $\omega$\,.
2d integrable field theories are realized by a reduction procedure.
In this formalism, the origin of the flatness condition is the bulk equation of motion of the 4d CS theory.
The choice of $\omega$ and the boundary condition determines the associated 2d theory.

\medskip

One may include two kinds of defects, order defects and disorder defects
(For the terminology, see \cite{CY}).
Disorder defects lead to various nonlinear sigma models and their integrable deformations\cite{DLMV,FSY1,FSY2,Fukushima:2021eni,Schmidtt:2019otc,Tian:2020ryu,Tian:2020meg,Hoare:2020mpv,Lacroix:2020flf,Costello:2020lpi,Bittleston:2020hfv}, and it is closely related to the affine Gaudin formalism\cite{Vicedo:2017cge,Gaudin,Gaiotto:2020fdr,Gaiotto:2020dhf}.
On the other hand, order defects give rise to the models with ultralocal Poisson structures such as the Zakharov-Mikhailov theory\cite{CSV} and the Faddeev-Reshetikhin model\cite{Fukushima:2020tqv}.
Although the sine-Gordon model has an ultralocal Lax pair, its prescription from the viewpoint of the 4d CS theory has not been clarified.

\medskip

Significant examples of 2d integrable field theories are non-abelian Toda field theories (NATFTs).
These include non-abelian extensions of sine-Gordon model and the Liouville theory, and the Lagrangian formulation of them is based on the Wess-Zumino-Witten (WZW) model plus a potential term \cite{Hollowood:1994vx,Bakas:1995bm,Fernandez-Pousa:1996aoa}.
It consists of a field which takes values in a non-abelian Lie group $G_{0}$\,, while the one in the standard Toda field theories takes values in a Cartan subgroup of a Lie group.
NATFTs may arise as the Pohlmeyer reduction of a certain class of sigma models\cite{Pohlmeyer:1975nb,Grigoriev:2007bu,Miramontes:2008wt}.
As a canonical example, the complex sine-Gordon model is obtained from the $R\times S^3$ sigma model, and soliton solutions are studied via this reduction\cite{Chen:2006gea,Okamura:2006zv}.

\medskip

The purpose of this paper is to derive the NATFTs from a 4d CS theory with two order defects.
We adopt $\omega$ with two simple poles and no zeros, and each of the order defects is located at each pole.
Since the defect terms are quadratic in the gauge field $A$\,, the on-shell action encounters an apparent divergence similar to the one for the free scalar in \cite{CY}.
This can be avoided by considering an appropriate regularization and a certain asymptotic boundary condition around the defects.
As a result, an anisotropy parameter $\beta$ can be identified with the distance between the two defects.
As examples, we can derive the (complex) sine-Gordon model and the Liouville theory.

\medskip

This paper is organized as follows.
We first give a brief review on the Lagrange formulation of NATFTs, and introduce three concrete examples in Section \ref{sec:NATFT}.
In Section \ref{sec:4dCS-NATFT}, we describe the details of the how to derive the NATFTs from a 4d CS theory.
Constraints on the gauge field and the gauge invariance are also discussed.
Section \ref{sec:conclusion} is devoted to conclusion and discussion. Appendix A explains how to realize the gauge 
choice utilized in our derivation.

\section{Non-abelian Toda field theory}\label{sec:NATFT}

In this section, we will briefly introduce the Lagrangian formulation of non-abelian Toda field theories (NATFTs) and present three concrete examples.

\subsection{The action of the NATFT}

Let us consider a NATFT on 2d Minkowski space $\cM$\,.
Given a compact Lie group $G$ with a Lie algebra $\mathfrak{g}$\,, the classical action of the NATFT is given by
\cite{Hollowood:1994vx}
\begin{align}
S_{\rm NATFT}[h]=&\,
\frac{1}{2\pi\beta^2}\left(S_{\rm WZW}[h]-\int_{\cM} d^2\sigma\; V(h)\right)\,,\qquad
V(h):=-m^2\langle\Lambda_{+}\,,\, h^{-1}\Lambda_{-} h\rangle\,,
\label{NATFT-action}\\
S_{\rm WZW}[h]:=&\,
-\frac{1}{2}\int_{\cM}d^2\sigma \left\langle h^{-1}\pa_{+}h\,,h^{-1}\pa_{-}h\right\rangle
-\int_{\cM\times[0,R_{x}]}I_{\rm WZ}[h]\,,\no\\
I_{\rm WZ}[h]:=&\,\frac{1}{3}\left\langle h^{-1}dh\,,h^{-1}dh\wedge h^{-1}dh\right\rangle\,,\no
\end{align}
where $h$ is a smooth function $\cM\to G$\,, and $\Lambda_{\pm}$ are generators of $\mathfrak{g}$\,.
The bracket $\langle \cdot\,,\cdot \rangle$ denotes an ad-invariant bilinear form.
The Lie algebra $\mathfrak{g}$ admits an action of involutive automorphism $\sigma$\,, which induces the following decomposition
\begin{align}
\mathfrak{g}=\mathfrak{g}_{0}\oplus \mathfrak{g}_{1}\,,\label{Z2-decom}
\end{align}
such that
\begin{align}
\sigma(\mathsf{x})=(-1)^p\mathsf{x}\,,\qquad {}^{\forall}\mathsf{x}\in\mathfrak{g}_{p}\,,\quad p=0,1\,.  \label{Z2-inv}
\end{align}
Here $\mathfrak{g}_{0}$ is a Lie subalgebra of $\mathfrak{g}$\,.
When a Lie group $G$ is compact, the associated NATFTs with a positive definite kinetic term are classified by the choice of $\Lambda_{\pm}$ as follows \cite{Fernandez-Pousa:1996aoa}:
\begin{align}\left\{\begin{array}{ll}
h\in G_{0}\,,\quad \Lambda_{\pm}\in\mathfrak{g}_{0}\quad & :\mbox{~a homogeneous sine-Gordon model}\\
h\in G_{0}\,,\quad \Lambda_{\pm}\in\mathfrak{g}_{1}\quad  & :\mbox{~a symmetric space sine-Gordon model}
\end{array}\right. \,. 
\end{align}
The Lie group $G_{0}$ is a subgroup of $G$ associated with $\mathfrak{g}_{0}$\,.
Note that the homogeneous sine-Gordon models can be described only by the elements of 
$\mathfrak{g}_{0}$\,.

\medskip

By taking a variation $\delta h=h\epsilon$\,, the action (\ref{NATFT-action}) varies as
\begin{align}
\delta S_{\rm NATFT}=\frac{1}{2\pi\beta^2}\bigg[
&\int_{\cM}d\sigma^{+}\wedge d\sigma^{-}\tr\Big[\epsilon\big(-\pa_{+}(h^{-1}\pa_{-}h)-\pa_{-}(h^{-1}\pa_{+}h)\big)\Big]\no\\
&-\int_{\cM}d\sigma^{+}\wedge d\sigma^{-}\tr(\epsilon[h^{-1}\pa_{+}h\,,\, h^{-1}\pa_{-}h])\no\\
&-2m^2\int_{\cM}d\sigma^{+}\wedge d\sigma^{-}\tr(\epsilon[\Lambda_{+}\,,\, h^{-1}\Lambda_{-} h])
\bigg]\,,
\end{align}
and thus we obtain the equation of motion (EOM)
\begin{align}
0=&\,\frac{1}{2}\Big(\pa_{+}(h^{-1}\pa_{-}h)+\pa_{-}(h^{-1}\pa_{+}h) + [h^{-1}\pa_{+}h\,,\, h^{-1}\pa_{-}h]
+2m^2[\Lambda_{+}\,,\, h^{-1}\Lambda_{-} h]\Big)\no\\
=&\,
\pa_{-}(h^{-1}\pa_{+}h)+m^2[\Lambda_{+}\,,\, h^{-1}\Lambda_{-} h]\,.  \label{NATFT-EOM-1}
\end{align}
This EOM is equivalent to
\begin{align}
0=\pa_{+}(h^{-1}\pa_{-}h)+m^2[h\Lambda_{+}h^{-1}\,,\Lambda_{-}]\,.  \label{NATFT-EOM-2}
\end{align}
These equations are can be expressed as the zero-curvature condition:
\begin{align}
0=&\,[\pa_{+}+\cL_{+}\,,\pa_{-}+ \cL_{-}]\,,\\
\cL_{+}=&\,h^{-1}\pa_{+}h+imw\Lambda_{+}\,,
\qquad\cL_{-}=\frac{im}{w}h^{-1}\Lambda_{-}h\,,
\label{NATFT-Lax}
\end{align}
where $w\in\mathbb{C}$ is the spectral parameter. 

\medskip 

Note here that EOMs (\ref{NATFT-EOM-1}) and (\ref{NATFT-EOM-2}) indicate that the potential induces the mass-gap along the $\operatorname{Im}(\operatorname{ad}\Lambda_{\pm})$-direction while flat directions still remain in general. 
If the adjoint actions $\operatorname{ad}_{\Lambda_{\pm}}$ on $\mathfrak{g}$ are diagonalizable and satisfy the condition
\begin{align}
\operatorname{Ker}(\operatorname{ad}_{\Lambda_{+}})\cap\mathfrak{g}_{0}=
\operatorname{Ker}(\operatorname{ad}_{\Lambda_{-}})\cap\mathfrak{g}_{0}\,,
\end{align}
then all of the flat directions of the potential can be gauged away. In order to see how this can be realized, let us couple the action $S_{\rm WZW}[h]$ with gauge fields $\mathcal{A}_{\pm}$ 
(which take values in $\operatorname{Ker}(\operatorname{ad}_{\Lambda_{\pm}})\cap\mathfrak{g}_{0}$) as the  gauged WZW action $S_{\rm WZW}[h,\cA_{\pm}]$ in (\ref{NATFT-action})\,.
Then the equation of motion takes the form
\begin{align}
0=\left[
\pa_{+}+h^{-1}\pa_{+}h + imw\Lambda_{+} +h^{-1}\cA_{+}h\;,\;
\pa_{-}+ \frac{im}{w}h^{-1}\Lambda_{-}h+\cA_{-}\right]\,. 
\label{NATFT-flat-w/gauge}
\end{align}
Taking variations with respect to $A_{\pm}$ lead to the constraints: 
\begin{align}\begin{split}
0=&\,P\left(h^{-1}\pa_{+}h+h^{-1}\cA_{+}h\right)-\cA_{+}\,,\\
0=&\,P\left(-\pa_{-}h\,h^{-1}+h\cA_{-}h^{-1}\right)-\cA_{-}\,,
\end{split}\label{gauge-constraint}
\end{align}
where $P$ denotes the projection onto $\operatorname{Ker}(\operatorname{ad}_{\Lambda_{\pm}})\cap\mathfrak{g}_{0}$\,.
The resulting theory is invariant under a gauge symmetry $h\mapsto \alpha h\alpha^{-1}$\,, where $\alpha$ is an element of the Lie group associated with $\operatorname{Ker}(\operatorname{ad}_{\Lambda_{\pm}})\cap\mathfrak{g}_{0}$\,. Note that the potential term $V(h)$ is invariant under this gauge transformation.

\subsection{Examples}\label{sec:example}

In the following, let us present three examples, i) the sine-Gordon model, ii) the Liouville theory, and iii) the complex sine-Gordon model.

\subsubsection*{i) sine-Gordon model}
The first example is the sine-Gordon model which corresponds to a symmetric space sine-Gordon model with $\mathfrak{g}=\mathfrak{su}(2)$ and $\mathfrak{g}_{0}=\mathfrak{u}(1)$\,. 

\medskip

Let us take the generators of $\mathfrak{su}(2)$ $T^{a}$ ($a=1,2,3$) as
\begin{align}
\left[T^{a},T^{b}\right]=\epsilon^{abc}T^{c}\,, \qquad
\left\langle T^{a}\,,T^{b}\right\rangle =-\frac{1}{2}\delta^{ab}\,,
\label{su(2)-gen}
\end{align}
where $\epsilon^{abc}$ is the anti-symmetric symbol with $\epsilon^{123}=1$\,. 
By choosing the $\mathfrak{u}(1)$ and the elements $\Lambda_{\pm}$
\begin{align}
h=\exp(\beta\phi T^3)\,,\qquad \Lambda_{+}=\Lambda_{-}=T^1\,, \qquad \phi\in\mathbb{R}
\label{parameter-SG}
\end{align} 
the action (\ref{NATFT-action}) and the Lax pair (\ref{NATFT-Lax}) are expressed as
\begin{align}
S_{\rm SG}[\phi]=&\,\frac{1}{4\pi}\int_{\cM}d^2\sigma\left( \frac{1}{2}\pa_{+}\phi\pa_{-}\phi - \frac{m^2}{\beta^2}\cos(\beta\phi) \right)\,,
\label{SG-action}\\
\begin{split}
\cL_{+}=&\,im wT^1+\beta\pa_{+}\phi T^3\,,\\
\cL_{-}=&\, \frac{im}{w}\Big(\cos(\beta\phi)T^{1} - \sin(\beta\phi)T^{2}\Big)\,.
\end{split}\end{align}
This is nothing but the sine-Gordon model.

\subsubsection*{ii) Liouville theory}
The Liouville field theory can be obtained by setting $\mathfrak{g}=\mathfrak{sl}(2,\mathbb{R})$ and $\mathfrak{g}_{0}=\mathbb{R}$\,.
The $\mathfrak{sl}(2,\mathbb{R})$ generators satisfy
\begin{align}\begin{split}
&\left[T^{1},T^{2}\right]=T^3\,,\quad \left[T^{0},T^{1}\right]=-T^2\,,\quad \left[T^{0},T^{2}\right]=T^1\,,\\
&\left\langle T^{0},T^{0}\right\rangle=-\frac{1}{2}\,,\quad \left\langle T^{1},T^{1}\right\rangle=\frac{1}{2}\,,\quad \left\langle T^{2},T^{2}\right\rangle=\frac{1}{2}\,,\quad
\mbox{otherwise}=0\,.
\end{split}\end{align}
Since $\mathfrak{sl}(2,\mathbb{R})$ and $\mathfrak{su}(2)$ have the same complexification $\mathfrak{sl}(2,\mathbb{R})^{\mathbb{C}}=\mathfrak{su}(2)^{\mathbb{C}}=\mathfrak{sl}(2,\mathbb{C})$\,,
the sine-Gordon model and the Liouville theory can be obtained from the same 4d CS theory. But the reality condition is different and the configuration of $\Lambda_{\pm}$ is also different.
To get a familiar form of the Liouville action, let us employ the following choice:  
\begin{align}
h=\exp(2\beta\phi T^1)\,,\quad
\Lambda_{+}=T^{2}+T^{0}\,,\quad \Lambda_{-}=T^{2}-T^{0}\,.
\end{align}
Then the resulting action and the associated Lax pair are given by 
\begin{align}
S_{\rm Liouville}[\phi]=&\,
\frac{1}{\pi}\int_{\cM}d^2\sigma\left(-\frac{1}{2}\pa_{+}\phi\pa_{-}\phi + \frac{m^2}{2\beta^2} e^{2\beta\phi} \right)\,,\\
\begin{split}
\cL_{+}=&\, imwT^{0} + 2\beta\pa_{+}\phi T^{1} + imw T^{2}\,,\\
\cL_{-}=&\, -\frac{im}{w}e^{2\beta\phi}T^{0} + \frac{im}{w}e^{2\beta\phi}T^{2}\,.
\end{split}\end{align}
This is the classical Liouville theory.

\subsubsection*{iii) complex sine-Gordon model}
The lowest rank example of non-trivial homogeneous sine-Gordon models is the complex sine-Gordon model, where the Lie algebra is given by $\mathfrak{g}_{0}=\mathfrak{su}(2)$\,.
In this case, a $\mathfrak{u}(1)$ gauge field can be turned on as in (\ref{NATFT-flat-w/gauge}).
Let us adopt the generators (\ref{su(2)-gen}) and take the following choice 
\begin{align}
\Lambda_{+}=T^3\,,\quad \Lambda_{-}=-T^{3}\,.
\end{align}
These obviously satisfy $\operatorname{Ker}(\operatorname{ad}_{\Lambda_{+}})\cap\mathfrak{g}_{0}=
\operatorname{Ker}(\operatorname{ad}_{\Lambda_{-}})\cap\mathfrak{g}_{0}=\operatorname{span}\{T^3\}$\,, and thus the gauge field $A_{\pm}$ takes values in $\operatorname{span}\{T^3\}$\,.
The dynamical variable $h$ is parametrized as
\begin{align}
h=\exp(\frac{\chi+\theta}{2}T^{3})\exp( (\phi-\pi) T^{1})\exp(\frac{\chi-\theta}{2}T^{3}) 
\qquad (\phi\,,\chi\,,\theta\in\mathbb{R})\,.
\end{align}
Since the parameter $\theta$ can be gauged out by the adjoint action, we work with the gauge $\theta=0$\,.
Then, the constraint (\ref{gauge-constraint}) can be solved by 
\begin{align}
\cA_{+}=\frac{\pa_{+}\chi}{2}\tan^2\left(\frac{\phi}{2}\right)\:T^{3}\,,\quad
\cA_{-}=-\frac{\pa_{-}\chi}{2}\tan^2\left(\frac{\phi}{2}\right)\:T^{3}\,,
\end{align}
which leads to the following expression of EOM:
\begin{align}
0=\pa_{+}\pa_{-}\psi + \frac{\bar{\psi}\pa_{+}\psi\pa_{-}\psi}{1-\psi\bar{\psi}} + m^2\psi(1-\psi\bar{\psi})\,,\quad
\psi:=\sin(\frac{\phi}{2})\exp(\frac{i\chi}{2})\,.
\end{align}
As a result, the  classical action can be expressed as
\begin{align}\begin{split}
S_{\rm CSG}[\psi]=&\, \frac{1}{2\pi\beta^2}\int d^2\sigma\left( \frac{\pa_{+}\bar{\psi}\pa_{-}\psi+\pa_{-}\bar{\psi}\pa_{+}\psi}{2(1-|\psi|^2)} - m^2 \Big(\frac{1}{2}-|\psi|^2\Big) \right)\\
=&\, \frac{1}{4\pi\beta^2}\int d^2\sigma\left( \frac{1}{2}\pa_{+}\phi\pa_{-}\phi 
-m^2\cos\phi + \frac{\tan^2(\phi/2)}{2}\pa_{+}\chi\pa_{-}\chi 
\right)\,.
\end{split}\end{align}
This is the action for the complex sine-Gordon model\cite{Okamura:2006zv}.
Note that when $\chi$ is constant, it reduces to the usual sine-Gordon model.

%
%

\section{NATFT from 4d CS theory}\label{sec:4dCS-NATFT} 

In this section, we will derive NATFTs from a 4d CS theory with two order defects.  

\subsection{4d CS theory with order defects}

First of all, let us introduce a 4d CS theory with two order defects. 
Although the gauge group $G$ was a Lie group in the previous section,  
we will work here with a complexified Lie group $G^{\mathbb{C}}$ and the associaed Lie algebra $\mathfrak{g}^{\mathbb{C}}$\,. 
The theory lives on a 4d manifold $\cM\times C$\,, 
where $\mathcal{M}$ is 2d Minkowski spacetime and $C:=\mathbb{C}^{\times}=\mathbb{C}\backslash\{0\}$\,.
The two defects are located at $z=\pm z_1$ on $C$\,. 
The gauge field $A$ is a smooth function $A:\cM\times C\to \mathfrak{g}^{\mathbb{C}}$ except for $z=\pm z_1$\,. In order to introduce the defects, we use two smooth functions 
$\cG_{(\pm)}:\cM\times C \to G^{\mathbb{C}}$\,.

\subsubsection*{Classical action} 

The classical action we consider is given by 
\begin{align}
S[\cG_{(\pm)},A]=&\, S_{\rm 4dCS}[A]+S_{\rm defect}[\cG_{(\pm)},A]\,,
\label{4dCS-action}\\
S_{\rm 4dCS}[A]:=&\, \frac{i}{4\pi }\int_{\cM\times C}\omega\wedge\left\langle A\,,\, dA+\frac{2}{3}A\wedge A\right\rangle\,,
\label{CS-term}
\\
S_{\rm defect}[\cG_{(\pm)},A]:=&\,
C_{(+)}\int_{\cM\times\{+ z_1\}} d\sigma^{+}\wedge d\sigma^{-}\left\langle \cG_{(+)}^{-1}D_{+}\cG_{(+)} \,,\, \cG_{(+)}^{-1}D_{-}\cG_{(+)} \right\rangle
\label{defect-term}\\
&+C_{(-)}\int_{\cM\times\{- z_1\}} d\sigma^{+}\wedge d\sigma^{-}\left\langle \cG_{(-)}^{-1}D_{+}\cG_{(-)} \,,\, \cG_{(-)}^{-1}D_{-}\cG_{(-)} \right\rangle\,,\no
\end{align}
where the covariant derivative $D_{\pm}$ is defined by
\begin{align}
D_{\pm}:=\pa_{\pm}+A_{\pm}\,.
\end{align}
The meromorphic 1-form $\omega$ here is taken as\footnote{This 1-form has no zeros and hence our analysis here cannot be based on disorder defects. } 
\begin{align}
\omega=\varphi(z)dz:=\frac{1}{(z-z_1)(z+z_1)}dz
\label{NATFT-twist}
\end{align}
and has two simple poles and no zeros. In total, the classical action contains three real constant parameters: 
$z_{1}\in \mathbb{R}$ and $C_{(\pm)}\in\mathbb{R}$\,. Note here that our defects are located at the positions of poles. 
This point is different from the treatment by Costello and Yamazaki \cite{CY}.  The reason of this choice will be clarified 
later.

\subsubsection*{Reality condition}

To ensure that the action (\ref{4dCS-action}) is real, some suitable conditions are necessary to be imposed.
The reality condition is described by introducing an involution $\mu_{\rm t}:C\to C$
\begin{align}
\mu_{\rm t}:z\mapsto \frac{|z_1|^2}{\bar{z}}\,,
\end{align}
and an involutive automorphism $\tau:\mathfrak{g}^{\mathbb{C}}\to\mathfrak{g}^{\mathbb{C}}$ which satisfies
\begin{align}
\overline{\left\langle\mathrm{x}\,,\mathrm{y}\right\rangle}=\left\langle\tau(\mathrm{x})\,,\tau(\mathrm{y})\right\rangle\,,\qquad \mathrm{x},\mathrm{y}\in\mathfrak{g}^{\mathbb{C}}\,.
\end{align}
The corresponding automorphism on the Lie group is denoted by $\tilde{\tau}:G^{\mathbb{C}}\to G^{\mathbb{C}}$\,.
The action (\ref{4dCS-action}) is real if the following condition holds:
\begin{align}
\mu_{\rm t}^{*}A=\tau A\,,\qquad \mu_{\rm t}^{*}\cG_{(\pm)}=\tilde{\tau}\cG_{(\pm)}\,.
\label{reality-condition}
\end{align}
Noting the fact $\overline{\omega}=\mu_{\rm t}^{*}\omega$ for $z_1\in\mathbb{R}$\,, one can confirm that 
\begin{equation}
\overline{S_{\rm 4dCS}[A]}=S_{\rm 4dCS}[A]\,, \qquad \overline{S_{\rm defect}[\cG_{(\pm)},A]}=S_{\rm defect}[\cG_{(\pm)},A]\,. 
\end{equation}
Due to the reality condition, the fields take values in the real Lie algebra and Lie group at fixed points of $\tilde{\tau}$\,,
and thus
\begin{align}
A|_{z=\pm z_{1}}\in \mathfrak{g}\,,\qquad \cG_{(\pm)}|_{z=\pm z_{1}}\in G\,.
\end{align}

\subsubsection*{Grading constraint}

Recall that the group-valued field $h$ of the NATFT is determined by considering the decomposition (\ref{Z2-decom}) of the Lie algebra $\mathfrak{g}$ with the $\mathbb{Z}_2$-graded involutive automorphism $\sigma$ in (\ref{Z2-inv}), as described in section \ref{sec:NATFT}.
This $\sigma$ can be identified in the context of the 4d CS theory by introducing an involution $\lambda: \cM\times C\to\cM\times C$ such that
\begin{align}
\lambda: (\sigma^{+},\sigma^{-},z)\mapsto \left(\sigma^{+},\sigma^{-},\frac{z_{1}^2}{z}\right)\,. 
\end{align}
Under the action of $\lambda$\,, the action (\ref{4dCS-action}) is invariant. 
Then by using this $\lambda$\,, the grading structure is induced as follows: 
\begin{align}
\lambda^{*}A=&\, \sigma(A)\,,\quad \lambda^{*}\cG_{(\pm)}= \tilde{\sigma}(\cG_{(\pm)})
 \qquad \mbox{:\,the grading constraint}\,, 
\label{lambda-constraint}
\end{align}
where $\sigma$ in (\ref{Z2-inv}) is naturally extended from $\mathfrak{g}$ to $\mathfrak{g}^{\mathbb{C}}$\,. 
The automorphism $\sigma$ also induces a map $\tilde{\sigma}:G^{\mathbb{C}}\to G^{\mathbb{C}}$\,.

\medskip 

Note that this grading structure requires that the $\mathfrak{g}_{1}$-component $A^{(1)}$  of gauge field $A$ should satisfy $A^{(1)} = - A^{(1)}$ 
at the fixed points of $\lambda$\,, i.e., $z=\pm z_{1}$\,. This equality is satisfied if and only if $A^{(1)}$ is $0$ or $\infty$\,. 
Here the possibility of $\infty$ is included because $A$ is not smooth at $z= \pm z_1$ in our setup.

\subsection{Equations of motion}

Let us derive the equations of motion (EOMs) for the classical action (\ref{defect-term}).
Taking a variation $\delta A$\,, we obtain the bulk EOMs
\begin{align}\begin{split}
F_{+-}=&\,0\,,\\
\delta A_{-}\cdot\varphi(z)F_{\bar{z}+}=&\, 
+2\pi i\,\delta A_{-}\cdot\sum_{x=\pm z_1}C_{(x)}\delta(z-x)D_{+}\cG_{(x)}\cdot \cG_{(x)}^{-1}\,, \\
\delta A_{+}\cdot\varphi(z)F_{\bar{z}-}=&\, 
-2\pi i\,\delta A_{+}\cdot\sum_{x=\pm z_1}C_{(x)}\delta(z-x)D_{-}\cG_{(x)}\cdot \cG_{(x)}^{-1}\,,
\end{split}\label{bulk-EOM-A}
\end{align}
and the boundary EOM 
\begin{align}
0=&\,
(\Res_{z=+z_1}\omega)\epsilon^{\alpha\beta}\left\langle A_{\alpha}|_{+z_1}\,,\delta A_{\beta}|_{+z_1}\right\rangle
+(\Res_{z=-z_1}\omega)\epsilon^{\alpha\beta}\left\langle A_{\alpha}|_{-z_1}\,,\delta A_{\beta}|_{-z_1}\right\rangle\no\\
=&\, \epsilon^{\alpha\beta}\big(\left\langle A_{\alpha}|_{+z_1}\,,\delta A_{\beta}|_{+z_1}\right\rangle-\left\langle A_{\alpha}|_{-z_1}\,,\delta A_{\beta}|_{-z_1}\right\rangle\big)\,.
\label{boundary-EOM}
\end{align}
In comparison to the disorder defect case, the bulk EOMs (\ref{bulk-EOM-A}) contain delta functions on the right-hand sides, 
while the boundary EOMs are the same as in the disorder defect case.  
This is a peculiar feature in the present case. 

\medskip 

Followng the Costello-Yamazaki scenario, in the first place, one needs to  determine a possible boundary condition satisfying the boundary EOMs (\ref{boundary-EOM}) so as to determine our setup. This process is the same as in the disorder defect case, 
but one should be careful for the bulk EOMs as well in the current case. 

\subsubsection*{Boundary conditions (from the boundary EOMs)} 

To fix our setup, it is necessary to determine a possible boundary condition satisfying the boundary EOMs (\ref{boundary-EOM})\,. 

\medskip 

In the following, let us suppose asymptotic behaviors of  $A$ around $z=\pm z_1$ as follows: 
\begin{align}
&A_{+}|_{z\to+z_1}=\cO(z-z_1)\,,\qquad A_{-}|_{z\to-z_1}=\cO(z+z_1)\,. \label{gauge-boundaryA}
\end{align}
This is not contradictory with the grading constraint (\ref{lambda-constraint})\,.  
Furthermore, to simplify our analysis, we suppose the following behaviors of $\cG_{(\pm)}$:  
\begin{align}
&\hspace{-10pt}\left.\left(
\pa_{+}\cG_{(+)}\right)\right|_{z\to+z_1}=\cO(z-z_1)\,,\quad
\left.\left(
\pa_{-}\cG_{(-)}\right)\right|_{z\to-z_1}=\cO(z+z_1)\,.
\label{gauge-boundary}
\end{align}
Under these boundary conditions, $A$ and $\cG_{(\pm)}$ can take the forms
\begin{align}\begin{split}
&A_{+}=(z-z_1)\mathscr{A}_{+}\,,\qquad A_{-}=(z+z_1)\mathscr{A}_{-}\,,\\
&\cG_{(+)}=(z- z_1)\tilde{\cg}_{(+)}(\sigma^+,\sigma^-)+\cg_{(+)}(\sigma^{-})\,,\\
&\cG_{(-)}=(z+ z_1)\tilde{\cg}_{(-)}(\sigma^+,\sigma^-)+\cg_{(-)}(\sigma^{+})\,,
\end{split}\end{align}
where $\mathscr{A}_{\pm}$\,, $\tilde{\cg}_{(\pm)}$ and $\cg_{(\pm)}$ are smooth functions with  $\mathscr{A}_{\pm}:\cM\times C\to \mathfrak{g}^{\mathbb{C}}$ and $\tilde{\cg}_{(\pm)}\,,\cg_{(\pm)}:\cM\to G^{\mathbb{C}}$\,, respectively.

\subsubsection*{Regularization (from the bulk EOMs)} 

Let us next discuss the bulk EOMs. 

\medskip 

Recall that the bulk EOMs (\ref{bulk-EOM-A}) contain delta functions on the right-hand sides. In order 
to satisfy these EOMs, delta functions should appear on the left-hand sides as well through the singular behavior of $A$ 
like\footnote{This singular behavior is not contradictory with the grading constraint (\ref{lambda-constraint})\,. } 
\begin{equation}
A _{\pm} \sim \frac{1}{z-x} \qquad (x = \pm z_1)\,. 
\end{equation}
This singularity comes from the existence of the defects. However, this singular behavior apprears on the right-hand sides thorugh 
$D_{\pm}$ and hence this contribution should be cancelled out by the poles of $\varphi(z)$\,. This is nothing but the reason that 
why the defect should be located at the poles of $\varphi(z)$\,. 

\medskip

For careful treatment, let us regularize the delta functions in (\ref{bulk-EOM-A}) and the singularity of $A$\,. 
As usual, the delta functions can be regularized by introducing smearing parameters $\alpha_{x}\in \mathbb{R}_{>0}$ ($x=\pm z_1$): 
\begin{align}\begin{split} 
 \int_{\cM\times\{x\}} &=\int_{\cM\times C}dz\wedge d\bar{z}\,\delta(z-x)\sim \, 
\int_{\cM\times C}dz\wedge d\bar{z}\,\frac{i e^{-|z-x|^2/\alpha_{x}}}{2\pi \alpha_{x}}\,.
\end{split}\label{delta-smear}
\end{align}
Then the singularity of $A$ can be regularized as 
\begin{align}
 A_{\pm }
&\sim\, \frac{1}{z-x}\left( 1- e^{-|z-x|^2/\alpha_{x}}\right)\,, 
\end{align}
so as to satisfy the relation $\pa_{\bar{z}}(1/z)=-2\pi i \delta(z)$\,.  
Note that the orientation of the complex plane is defined as
\begin{align}
1=\int_{C}dz\wedge d\bar{z}\,\delta(z)=-2\pi i \int_{C}dz\wedge d\bar{z}\;\pa_{\bar{z}}\frac{1}{z}=2\pi i\int_{C}d\bar{z}\wedge\pa_{\bar{z}}\left(\frac{dz}{z}\right)\,.
\end{align}
Unless otherwise noted, these regularizations are used. 

\subsection{Gauge invariance}\label{sec:gauge-inv}

In this subsection, let us discuss the gauge invariance of the action (\ref{4dCS-action})\,. Actually, the gauge invariance is not always ensured for arbitrary boundary conditions at $z=\pm z_1$ on the gauge field $A$ and the defect fields $\cG_{\pm}$\,. 
In the following, we will show that the boundary condition (\ref{gauge-boundaryA}) enjoys the gauge invariance. 

\medskip 

Under the boundary condition (\ref{gauge-boundaryA})\,, the gauge transformation 
\begin{align}
A\mapsto A^{u}:=uAu^{-1}-duu^{-1}\,,\qquad \cG_{(\pm)}\mapsto \cG_{(\pm)}^{u}:=u\, \cG_{(\pm)}
\label{gauge-trans}
\end{align}
is restricted because the gauge function $u$ $(u:\cM\times C\to G^{\mathbb{C}})$ needs to satisfy 
\begin{align}
\lambda^{*}u= \tilde{\sigma}(u)\,. \label{u-condi}
\end{align}
Note here that 
from the condition (\ref{u-condi}), $u$ at the defects can always be parametrized as 
\begin{align}
u|_{\pm z_1}=\exp(\alpha_a T^{a})\in G_0^{\mathbb{C}} \qquad (\alpha_a:\cM
\to\mathbb{C})\,, 
\label{complex}
\end{align}
where $T^a$ are the generators of $\mathfrak{g}_{0}^{\mathbb{C}}$\,.
Furthermore, the boundary condition (\ref{gauge-boundaryA}) restricts the gauge parameter $u$ as
\begin{align}
\pa_{+}u|_{+z_1}=0\,,\qquad \pa_{-}u|_{-z_1}=0\,.
\label{u-derivative}
\end{align}
Then the reality condition (\ref{reality-condition}) restricts the condition (\ref{complex}) to be real:
\begin{align}
u|_{\pm z_{1}}:\cM \to G_0\,,\qquad 
u|_{\pm z_1}=\exp(\alpha_a T^{a}) \qquad (\alpha: \cM \to\mathbb{R})\,.
\label{u-parameter}
\end{align} 

For later convenience, it is helpful to make a comment on the transformations which do not satisfy (\ref{u-derivative}) and (\ref{u-parameter})\,. In our terminology, these are referred to as {\it formal gauge transformations} since such transformations are not symmetry of the action in general. These transformations are utlized in the next subsection. 

\medskip 

Let us show the gauge invariance of th action (\ref{4dCS-action})\,. 
By construction, the defect terms (\ref{defect-term}) are invariant under the gauge transformation (\ref{gauge-trans})\,. 
Hence, we shall focus on the CS part (\ref{CS-term}) in the following. 
Under the gauge transformation (\ref{gauge-trans}), the CS part (\ref{CS-term}) varies as
\begin{align}
S_{\rm 4dCS}[A^u]=S_{\rm 4dCS}[A] + \frac{i}{4\pi}\int_{\cM\times C} \omega \wedge I_{\rm WZ}[u] + \frac{i}{4\pi}\int_{\cM\times C} \omega\wedge d\langle u^{-1}du,A\rangle\,.
\label{gauge variation}
\end{align}
For the gauge invariance, the second and third terms of (\ref{gauge variation}) should vanish.  

\medskip

To show this, it is convenient to utilize the following formula:
\begin{align}
\int_{\cM\times C}\omega\wedge d\eta = 2\pi i\sum_{x\in\mathfrak{p}} k_{x}\int_{\cM\times \{x\}}\eta\,,
\label{pole-localize}
\end{align}
where $\eta$ is an arbitrary 2-form-valued smooth function and $k_{x}$ denotes the residues of $\omega$\,. 
In the case of the twist function (\ref{NATFT-twist}), we have $k_{+z_1}=-k_{-z_1}=1/(2z_1)$\,. 

Then the formula (\ref{pole-localize}) can be shown as
\begin{align}
\int_{\cM\times C}\omega\wedge d\eta=&\,\int_{\cM\times C}\omega\wedge\big(d_{\cM}+d\bar{z}\,\pa_{\bar{z}}\big) \eta
= \int_{\cM\times C}\omega\wedge d\bar{z}\,\pa_{\bar{z}}\eta -\int_{\cM\times C}d_{\cM}\big(\omega\wedge\eta\big)\no\\
=&\,2\pi i\sum_{x\in\mathfrak{p}} k_{x}\int_{\cM\times \{x\}}\eta\,,
\end{align}
where $d_{\cM}:=d\tau\pa_{\tau}+d\sigma\pa_{\sigma}$\,. 
In the second line, we utilized the Cauchy-Pompeiu formula
\begin{align}
f(x)=\frac{1}{2 \pi i }\oint_{\pa C}\frac{f(z,\bar{z})dz}{z-x} +\frac{1}{2\pi i}\int_{C}\frac{dz\wedge d\bar{z}}{z-x}\pa_{\bar{z}}f(z,\bar{z})\,,\qquad
f\in C^{1}(C)\,.
\end{align}
Note here that the manifold $C=\mathbb{C}^{\times}$ has boundaries at $z=0\,,\infty$\,, and the first term of the right-hand side equals to zero because of the regularity of $f(z,\bar{z})$\,\footnote{
The contour integral around $z=\infty$ is evaluated as follows:
we define a local coordinate around $\infty$ by $\xi_{\infty}:=1/z$\,.
In the case of the twist function (\ref{NATFT-twist}), the corresponding integral is given by $\oint f(z,\bar{z})dz/z^2=-\oint f(\xi_{\infty}^{-1},\overline{\xi_{\infty}}^{-1})d\xi_{\infty}$ at $z\to \infty$\,.
This integral vanishes if $f(\xi_{\infty}^{-1},\overline{\xi_{\infty}}^{-1})\to0$ at the $\xi_{\infty}\to0$ limit.
}. 

\medskip

It is also helpful to use the formula
\begin{align}
\int_{\cM\times C}\omega\wedge I_{\rm WZ}[g]=-2\pi i\sum_{x\in\mathfrak{p}}k_{x}\int_{\cM\times[0,R_{x}]}I_{\rm WZ}[\bar{g}]\,.
\label{WZ-homotopy}
\end{align}
Here $\bar{g}:\cM\times[0,R_{x}]\to G^{\mathbb{C}}$ is a lazy homotopy such that $\bar{g}(\sigma^{\pm},0)=g(\sigma^{\pm},x)$ and $\bar{g}(\sigma^{\pm},R_{x})=e$\,, where $e$ is a constant map to the identity element of $G$\,.
 Note that the integral (\ref{WZ-homotopy}) does not depend on the choice of $\bar{g}$\,.
The expression (\ref{WZ-homotopy}) corresponds to Proposition 2.10 in \cite{BSV}.

\medskip 

By the use of the formulae (\ref{pole-localize}) and (\ref{WZ-homotopy})\,, let us evaluate the second and third terms in 
(\ref{gauge variation})\,. First of all, the third term of (\ref{gauge variation}) can be evaluated by taking $\eta = \langle u^{-1}du, A  \rangle$ 
in the formula  (\ref{pole-localize})\,. Then by using the boundary conditions (\ref{gauge-boundaryA}) and (\ref{u-derivative}) of $A$ and $u$ respectively\,, 
the third term vanishes. 

\medskip 

To evaluate the second term of (\ref{gauge variation})\,, we take $g = u$ in the formula (\ref{WZ-homotopy}) and 
the value of the right-hand side depends only on $u(\sigma^{\pm},x)$\,. Then due to the boundary condition (\ref{u-derivative}),  
we can always take $\pa_{\pm}u=0$ inside $\cM\times [0,R_{\pm z_1}]$\,.

\medskip 

After all, we obtain $S_{\rm 4dCS}[A^{u}]=S_{\rm 4dCS}[A]$\,, and then the action (\ref{CS-term}) is gauge-invariant.

\subsection{Lax form}\label{sec:Lax-form}

Next, let us introduce the Lax form associated with the action (\ref{4dCS-action}).

\medskip


The Lax form $\mathcal{L}$ is introduced by using a formal gauge transformation with $\hat{g}:\cM\times C\to G^{\mathbb{C}}$ as
\begin{align}
A=-d\hat{g}\hat{g}^{-1}+\hat{g}\cL\hat{g}^{-1}\,,\qquad \cG_{(\pm)}=\hat{g} g_{(\pm)}\,.
\label{formal-Lax}
\end{align}
Here we take a gauge $\cL_{\bar{z}} =0$ 
by taking the following gauge field configuration 
\begin{align}
A_{\bar{z}}=-(\pa_{\bar{z}}\hat{g})\,\hat{g}^{-1}\,.
\end{align} 
That is, $\mathcal{L}$ takes the form 
\begin{eqnarray}
\mathcal{L} = \mathcal{L}_{\tau} d\tau + \mathcal{L}_{\sigma} d\sigma = \mathcal{L}_+ d\tau^+ + \mathcal{L}_- d\sigma^-\,.  
\end{eqnarray}
The choice of $\hat{g}$ and $g_{(\pm)}$ has an ambiguity under the transformation
\begin{align}
\cL\mapsto \ch^{-1}d\ch+\ch^{-1}\cL\, \ch\,,\quad\hat{g}\mapsto \hat{g}\ch\,,\quad g_{(\pm)}\mapsto \ch^{-1}g_{(\pm)}\,,
\label{2d-gauge}
\end{align}
where $\ch$ is a smooth function of $\cM$ to $\mathfrak{g}^{\mathbb{C}}$\,.
In contrast to the gauge symmetry (\ref{gauge-trans}), the gauge field $A$ and the defect fields $\cG_{(\pm)}$ are invariant under the transformation (\ref{2d-gauge}), and hence the action (\ref{4dCS-action}) is also invariant.
In the following, we will call (\ref{2d-gauge}) the {\it 2d gauge transformation}.

\medskip

In terms of the Lax form $\cL_{\pm}$ and the defect fields $g_{(\pm)}$\,, the bulk EOMs in (\ref{bulk-EOM-A}) read
\begin{align}
0=&\,\pa_{+}\cL_{-}-\pa_{-}\cL_{+} + [\cL_{+},\cL_{-}]\,,
\label{on-shell-flatness}\\
\begin{split}
\frac{1}{z^2-z_{1}^2}\delta A_{-}\pa_{\bar{z}}\cL_{+}=&\, 
+2\pi i\,\delta A_{-}\sum_{x=\pm z_1}C_{(x)}\delta(z-x)(\pa_{+}+\cL_{+})g_{(x)}\cdot g_{(x)}^{-1}\,, \\
\frac{1}{z^2-z_{1}^2}\delta A_{+}\pa_{\bar{z}}\cL_{-}=&\, 
-2\pi i\,\delta A_{+}\sum_{x=\pm z_1}C_{(x)}\delta(z-x)(\pa_{-}+\cL_{-})g_{(x)}\cdot g_{(x)}^{-1}\,.
\end{split}
\label{Lax-EOM}
\end{align}
The $\bar{z}$-derivative of the Lax form $\pa_{\bar{z}}\cL$ vanishes at the points other than $z=\pm z_1$\,. The Lax form is thus a meromorphic 1-form with the poles at $z=\pm z_{1}$\,.
Then, a possible ansatz for $\cL$ is given by
\begin{align}\begin{split}
\cL_{+}=&\,\frac{U_{1,+}z+U_{0,+}z_{1}}{z+z_1}\,,\\
\cL_{-}=&\,\frac{U_{1,-}z+U_{0,-}z_{1}}{z-z_1}\,,
\end{split}\end{align}
where $U_{1,\pm}$, $U_{0,\pm}$ are smooth functions $\cM\to \mathfrak{g}^{\mathbb{C}}$\,.

\medskip

Recalling the relation (\ref{formal-Lax}), sufficient conditions for the grading constraint (\ref{lambda-constraint}) are given by 
\begin{align}
\lambda^{*}\cL=\sigma(\cL)\,,\qquad
\lambda^{*}\hat{g}=\tilde{\sigma}(\hat{g})\,, \label{grading-Lax}
\end{align}
because the map $\sigma$ is automorphic.
More concretely, the constraint (\ref{grading-Lax}) leads to the following expressions: 
\begin{align}
&\left\{\begin{array}{l}\vspace{5pt}
\sigma(\cL_{+})= \lambda^{*}\cL_{+}={\displaystyle \frac{U_{1,+}z_{1}^2/z+U_{0,+}z_{1}}{z_{1}^2/z+z_{1}}
=\frac{U_{0,+}z+U_{1,+}z_{1}}{z+z_{1}}}\,,\\ 
\sigma(\cL_{-})= \lambda^{*}\cL_{-}= {\displaystyle \frac{U_{1,-}z_{1}^2/z+U_{0,-}z_{1}}{z_{1}^2/z-z_{1}}
=-\frac{U_{0,-}z+U_{1,-}z_{1}}{z-z_{1}}}\,,
\end{array}\right.
\no\\
\Leftrightarrow&\qquad
U_{0,+}=\sigma(U_{1,+})\,,\quad U_{0,-}=-\sigma(U_{1,-})\,.
\end{align}
Then the Lax form can be rewritten as 
\begin{align}\begin{split}
\cL_{+}=&\, \frac{U_{0,+}z+\sigma(U_{0,+})z_{1}}{z+z_{1}}=\frac{z-z_1}{z+z_1}\widetilde{V}_{+} +V_{+}\,,\\
\cL_{-}=&\, \frac{U_{0,-}z+\sigma(U_{0,-})z_{1}}{z-z_{1}}=\frac{z+z_1}{z-z_1}\widetilde{V}_{-} +V_{-}\,,\\
\end{split}\label{Lax-ansatz}
\\
&V_{\pm}:=\frac{U_{0,\pm}+\sigma(U_{0,\pm})}{2}\,,\quad \widetilde{V}_{\pm}:=\frac{U_{0,\pm}-\sigma(U_{0,\pm})}{2}\,.
\end{align}
We can see that $V_{\pm}\in\mathfrak{g}_{0}$ and $\widetilde{V}_{\pm}\in\mathfrak{g}_{1}$\,.
Moreover, the reality condition (\ref{reality-condition}) constrains $V_{\pm}\,,\widetilde{V}_{\pm}$  as
\begin{align}
\overline{\widetilde{V}_{\pm}}=-\widetilde{V}_{\pm}\,,\qquad \overline{V_{\pm}}=V_{\pm}\,.
\label{reality-U}
\end{align}
Here we have taken a regularization similar to (\ref{delta-smear}) like 
\begin{equation}
\frac{1}{z-x} \sim \frac{1}{z -x}\, (1-\exp(-|z- x|^2/\alpha_{x}))\,.
\end{equation}

\medskip

Utilizing the gauge transformations, one of natural parametrizations is given by
\begin{align}
\hat{g}|_{z_1}=h\,,\quad \hat{g}|_{-z_1}=1\,,\quad g_{(+)}|_{z_1}=h^{-1}\,,\quad g_{(-)}|_{-z_1}=1\,.
\label{gauge-choice}
\end{align}
For the detail of realizing this gauge, see Appendix \ref{app:gauge-choice}.

\medskip

By substituting (\ref{Lax-ansatz}) and (\ref{gauge-choice}), the EOMs (\ref{Lax-EOM}) read
\begin{align}\begin{split}
\epsilon_{-}\cdot\delta(z+z_1) \widetilde{V}_{+}=
&- C_{(+)}\epsilon_{-}\cdot\delta(z-z_1)\left( -(z+z_1)h^{-1}\pa_{+}h + (z-z_1)\widetilde{V}_{+} +  (z+z_1)V_{+} \right)\\
&- C_{(-)}\epsilon_{-}\cdot\delta(z+z_1)\left( \frac{1}{2}(z-z_1)\widetilde{V}_{+} +  (z+z_1)V_{+} \right)\,,\\
\epsilon_{+}\cdot\delta(z-z_1) \widetilde{V}_{-}=
&+C_{(+)}\epsilon_{+}\cdot\delta(z-z_1)\left( -(z-z_1)h^{-1}\pa_{-}h + \frac{1}{2}(z+z_1)\widetilde{V}_{-} + (z-z_1)V_{-} \right)\\
&+C_{(-)}\epsilon_{+}\cdot\delta(z+z_1)\left( (z+z_1)\widetilde{V}_{-} +  (z-z_1)V_{-} \right)\,,
\end{split}\label{V-EOM-1}
\end{align}
where the variation $\delta A_{\pm}$ is expressed as
\begin{align}
\delta A_{+}=(z-z_1)\epsilon_{+} \,, \quad 
\delta A_{-}=(z+z_1)\epsilon_{-}\,,\qquad \epsilon_{\pm}|_{z\to\pm z_{1}}=\cO(1)\,.
\end{align}
In the computation of (\ref{V-EOM-1}), 1/2 factors have appeared because 
the regularization factor can be evaluated as 
\begin{align}
\frac{ie^{-|z-x|^2/\alpha_{x}}}{2\pi \alpha_{x}}\cdot\big(1-e^{-|z-x|^2/\alpha_{x}}\big)
=\frac{ie^{-|z-x|^2/\alpha_{x}}}{2\pi \alpha_{x}} - \frac{1}{2}\frac{ie^{-|z-x|^2/(\alpha_{x}/2)}}{2\pi (\alpha_{x}/2)}
\sim \frac{1}{2}\delta(z-x)\,.
\label{factor-1/2}
\end{align}
The bulk EOMs thus read
\begin{align}\begin{split}
\delta(z+z_1)\widetilde{V}_{+}=& +2z_1 C_{(+)}\delta(z-z_1)\big(-h^{-1}\pa_{+}h +V_{+}\big) - z_1C_{(-)}\delta(z+z_1)\widetilde{V}_{+}\,,\\
\delta(z-z_1)\widetilde{V}_{-}=& -z_1 C_{(+)}\delta(z-z_1)\widetilde{V}_{-} + 2z_1C_{(-)}\delta(z+z_1)V_{-}\,.
\end{split}\label{V-EOM-2}
\end{align}
To avoid a trivial solution $\widetilde{V}_{\pm}=0$\,, the constants should be taken as
\begin{align}
C_{(+)}=C_{(-)}=\frac{1}{z_1}\,,
\end{align}
and then the solution to (\ref{V-EOM-2}) is obtained as
\begin{align}
V_{+}=h^{-1}\pa_{+}h\,,\qquad V_{-}=0\,.
\end{align}

\medskip

Let us consider a variation $\delta \cG_{(\pm)}=\cG_{(\pm)}\epsilon_{(\pm)}$\,, such that $\pa_{\pm}\epsilon_{(\pm)}|_{z\to\pm z_1}=\cO(z\mp z_1)$\,. Then, due to the boundary condition (\ref{gauge-boundary}), $\epsilon_{(\pm)}$ can be rewritten  as
\begin{align}
\epsilon_{(\pm)}=(z\mp z_1)\tilde{\varepsilon}_{(\pm)}(\sigma^+,\sigma^-)+\varepsilon_{(\pm)}(\sigma^{\mp})\,,\qquad
\tilde{\varepsilon}|_{z\to\pm z_1}=\cO(1)\,. \label{expansion}
\end{align}
By using (\ref{expansion})\,, the EOMs with respect to $\delta{\cG}_{(\pm)}$ are given by
\begin{align}\begin{split}
0=&\,
(z-z_1)\tilde{\varepsilon}_{(+)}\pa_{+}\left( h (\pa_{-} +  \frac{1}{2}\frac{z+z_1}{z-z_1}\widetilde{V}_{-} )h^{-1} \right)\bigg|_{z\to +z_1}\\
&+\varepsilon_{(+)}\pa_{-}\left( h(\pa_{+} + h^{-1}\pa_{+}h + \frac{z-z_1}{z+z_1}\widetilde{V}_{+})h^{-1}\right)\bigg|_{z\to +z_1}\,,
\\
0=&\,\varepsilon_{(-)}\pa_{+}\left(   \frac{z+z_1}{z-z_1}\widetilde{V}_{-}  \right)\bigg|_{z\to -z_1}
+(z+z_1)\tilde{\varepsilon}_{(-)}\pa_{-}\left(  h^{-1}\pa_{+}h + \frac{1}{2}\frac{z-z_1}{z+z_1}\widetilde{V}_{+}\right)\bigg|_{z\to -z_1}\,,
\end{split}\label{EOM-defect-NATFT}
\end{align}
where the $1/2$ factors due to (\ref{factor-1/2}) have appeared again.
In the $\alpha_{x}\to0$ limit, the relevant terms are expressed as
\begin{align}\begin{split}
&\pa_{+}\left(h \widetilde{V}_{-} h^{-1}\right)=0\,,\qquad \pa_{-}\widetilde{V}_{+}=0\,,\\
\Rightarrow\quad & \widetilde{V}_{-}=im\, h^{-1}\Lambda_{-}h\,,\qquad \widetilde{V}_{+}=im\,\Lambda_{+}\,,
\end{split}\end{align}
where we introduced smooth functions $\Lambda_{\pm}:\cM\to \mathfrak{g}_{0}$ such that $\pa_{\mp}\Lambda_{\pm}=0$\,.
After all, the solution to EOMs (\ref{Lax-EOM}) and (\ref{EOM-defect-NATFT}) are given by
\begin{align}\begin{split}
\cL_{+}=&\,h^{-1}\pa_{+}h + im\frac{z-z_1}{z+z_1}\Lambda_{+}\,,\\
\cL_{-}=&\,im\frac{z+z_1}{z-z_1}h^{-1}\Lambda_{-}h\,.
\end{split}\label{Lax-solution}
\end{align}
This is exactly the Lax pair for the NATFTs (\ref{NATFT-Lax}) with $w=(z-z_1)/(z+z_1)$\,.
Note that $\Lambda_{\pm}$ may depend on $\sigma^{\pm}$ here, in comparison to the usual case where these are constants.

\medskip

At a glance, the Lax pair (\ref{Lax-solution}) is written with the rational form of $z$ and this seems to indicate that the system 
should be of rational type. However, this is not written 
in terms of $z-z_1$ only. In fact, the Lax pair (\ref{Lax-solution}) depends only on $z/z_1$ 
and  
we can see that the system is of trigonometric type as follows.
To make the periodicity explicit, one can define a map from the cylinder $\mathbb{C}/\mathbb{Z}$ to $\mathbb{C}^{\times}:=\mathbb{C}P^1\backslash\{ 0,\infty\}$ 
\begin{align}
f:\widetilde{z}\mapsto z=e^{\alpha\widetilde{z}}\,.
\end{align}
In terms of $\tilde{z} \in \mathbb{C}/\mathbb{Z}$\,, the expression (\ref{Lax-solution}) is given by
\begin{align}
f^{*}\cL_{+}=&\,h^{-1}\pa_{+}h+im \coth(\frac{\alpha\widetilde{z}-\alpha\widetilde{z}_1-i\pi}{2})\Lambda_{+}
=h^{-1}\pa_{+}h+im \tanh(\frac{\alpha\widetilde{z}-\alpha\widetilde{z}_1}{2})\Lambda_{+}\,,
\no\\
f^{*}\cL_{-}=&\,im\coth(\frac{\alpha\widetilde{z}-\alpha\widetilde{z}_1}{2})h^{-1}\Lambda_{-}h\,,
\end{align}
where $\tilde{z}_1$ is defined through $z_1=:\exp(\alpha\widetilde{z}_1)$\,. Thus, we see that the system is of trigonometric. 

\medskip 

A rational limit is realized by rescaling $\Lambda_{\pm}$ as $\Lambda_{\pm}=\alpha^{\mp 1}\widetilde{\Lambda}_{\pm}$ and 
then taking the limit $\alpha\to0$\,. The resulting Lax pair takes the rational form as follows: 
\begin{align}
f^{*}\cL_{+}=h^{-1}\pa_{+}h+im\frac{\widetilde{z}-\widetilde{z}_1}{2}\widetilde{\Lambda}_{+}\,,& \quad
f^{*}\cL_{-}=im\frac{2}{\widetilde{z}-\widetilde{z}_1}h^{-1}\widetilde{\Lambda}_{-}h\,,\\
z_{1}=1\,.
\end{align}

\subsection{Reduction to 2d action}

Now let us reduce the action (\ref{4dCS-action}) to the associated 2d action.

\medskip

For this purpose, we first rewrite the 4d action (\ref{4dCS-action}) in terms of the Lax form.
Under the formal gauge transformation (\ref{formal-Lax}), the CS part (\ref{CS-term}) varies according to the formula (\ref{gauge variation}) with $u=\hat{g}$, and the 4d action (\ref{4dCS-action}) can be rewritten as
\begin{align}
S[\cG_{(\pm)},A]=&\, \frac{i}{4\pi}\int_{\cM\times C} \omega \wedge I_{\rm WZ}[\hat{g}] + \frac{i}{4\pi}\int_{\cM\times C} \omega\wedge d\tr\left(\hat{g}^{-1}d\hat{g}\,\cL\right)\no\\
&\,+\frac{i}{4\pi}\int_{\cM\times C}\omega\wedge \tr(\cL\wedge d \cL)+S_{\rm defect}[\cL,h]\,,\label{Lax-4dCS}\\
S_{\rm defect}[\cL, h]=&\, 
\int_{\cM\times \{+z_1\}}\frac{d\sigma^{+}\wedge d\sigma^{-}}{z_1}\tr(h(\pa_{+}+\cL_{+})h^{-1}\cdot h (\pa_{-}+\cL_{-})h^{-1})\no\\
&+\int_{\cM\times \{-z_1\}}\frac{d\sigma^{+}\wedge d\sigma^{-}}{z_1}\tr(\cL_{+}\cL_{-})\,.\label{Lax-defect}
\end{align}
The first and second terms in the first line of (\ref{Lax-4dCS}) are already evaluated at (\ref{WZ-homotopy}) and (\ref{pole-localize}) with $\eta= \frac{i}{4\pi}\tr\left(\hat{g}^{-1}d\hat{g}\,\cL\right)$, respectively.
Hence, the action (\ref{Lax-4dCS}) is simplified to
\begin{align}
S_{\rm 2d}[h;\Lambda_{\pm}]=&\,
\sum_{x=\pm z_1}\frac{1}{2}\int_{\cM}\tr\Big((\Res_{x}\varphi\cL)\wedge g_{x}^{-1}dg_{x}\Big)
- \frac{1}{2}\sum_{x=\pm z_{1}}\int_{\cM\times[0,R_{x}]}(\Res_{x}\omega)\wedge I_{\rm WZ}[\bar{g}_{x}]\no\\
&+\frac{i}{4\pi}\int_{\cM\times C}\omega\wedge \tr(\cL\wedge d \cL) + S_{\rm defect}[\cL, h]\,,
\label{2d-formula}
\end{align}
where $g_{x}:=g|_{z=x}$\,, and $\bar{g}_x:\cM\times[0,R_{x}]\to G^{\mathbb{C}}$ is a smooth map such that $\bar{g}_x(\sigma^{\pm},R_x)=g_x(\sigma^{\pm},x)$ and $\bar{g}_x(\sigma^{\pm},0)=e$\,.
This action is almost a 2d object, but the first term in the second line of (\ref{2d-formula}) is still a 4d integral.

\medskip

Next, we substitute the explicit expression (\ref{Lax-solution}) of the Lax form $\cL$ into (\ref{2d-formula}).
The first line of (\ref{2d-formula}) is reduced to the action of the PCM with WZ term
\begin{align}
\sum_{x=\pm z_1}&\frac{1}{2}\int_{\cM}\tr\Big((\Res_{x}\varphi\cL)\wedge g_{x}^{-1}dg_{x}\Big)
- \frac{1}{2}\sum_{x=\pm z_{1}}(\Res_{x}\omega)\int_{\cM\times[0,R_{x}]}I_{\rm WZ}[\bar{g}_{x}]\no\\
=&\,
\frac{1}{4z_1}\int_{\cM}d\sigma^{+}\wedge d\sigma^{-}\tr(h^{-1}\pa_{+}h\,h^{-1}\pa_{-}h)
- \frac{1}{4z_1}\int_{\cM\times[0,R_{x}]}I_{\rm WZ}[\bar{h}]\,.
\end{align}
The first term in the second line of (\ref{2d-formula}) is evaluated as
\begin{align}
&\frac{i}{4\pi}\int_{\cM\times C}\omega\wedge \tr(\cL\wedge d \cL)\no\\
&=\, \frac{i}{4\pi}\int_{\cM\times C} dz\wedge d\sigma^{+}\wedge d\bar{z}\wedge d\sigma^{-}\left[m^2 \frac{1}{z+z_1}2\pi i\delta(z-z_1)\tr(\Lambda_{+}h^{-1}\Lambda_{-}h)\right]\no\\
&\quad+ \frac{i}{4\pi}\int_{\cM\times C} dz\wedge d\sigma^{-}\wedge d\bar{z}\wedge d\sigma^{+}\left[m^2 \frac{1}{z-z_1}2\pi i\delta(z+z_1)\tr(h^{-1}\Lambda_{-}h\Lambda_{+})\right]\no\\
&=\,
\frac{m^2}{2z_1}\int_{\cM}d\sigma^{+}\wedge d\sigma^{-}  \tr(\Lambda_{+}h^{-1}\Lambda_{-}h)\,.
\end{align}
Thus, the 4d integral is reduced to a 2d one.
Finally, the defect action (\ref{Lax-defect}) becomes
\begin{align}
S_{\rm defect}[\cL, h]
=&\,
-\int_{\cM}d\sigma^{+}\wedge d\sigma^{-}\frac{m^2}{z_1}\tr(\Lambda_{+}h^{-1}\Lambda_{-} h)\,.
\end{align}
After all, the resulting 2d action is obtained as
\begin{align}
S_{\rm 2d}[h;\Lambda_{\pm}]=&\,
\frac{1}{4z_1}\int_{\cM}d\sigma^{+}\wedge d\sigma^{-}\tr(h^{-1}\pa_{+}h\,h^{-1}\pa_{-}h)
- \frac{1}{4z_1}\int_{\cM\times[0,R_{x}]}I_{\rm WZ}[\bar{h}]\no\\
&-\frac{m^2}{2z_1}\int_{\cM}d\sigma^{+}\wedge d\sigma^{-}  \tr(\Lambda_{+}h^{-1}\Lambda_{-}h)\,. \label{2daction-4d}
\end{align}
The expression (\ref{2daction-4d}) precisely agrees with the action (\ref{NATFT-action}) of the NATFT
by relating the parameters $\beta$ and $z_1$ in the relation
\begin{align}
\frac{1}{2\pi\beta^2}=\frac{1}{4z_1}\,.
\end{align}

\medskip

In the discussion so far, we have fixed the coefficients of the the defects for $C_{(+)}=C_{(-)}=1/z_1$ so that the bulk EOMs (\ref{V-EOM-2}) have a nontrivial solution.
Otherwise, the Lax pair becomes $\cL_{+}=h^{-1}\pa_{+}h$ and $\cL_{-}=0$\,, and the 2d action is just the one of the conformal WZW model.

\subsubsection*{Another gauge choice for the sine-Gordon model}
As discussed in section \ref{sec:example}, the sine-Gordon model corresponds to $\mathfrak{g}_{0}=\mathfrak{u}(1)$\,.
In the case where $\mathfrak{g}_{0}$ is a direct product of $\mathfrak{u}(1)$\,, we can choose a different gauge than (\ref{gauge-choice}).
The fields $\hat{g}$ and $g_{(\pm)}$ at each pole are expressed as
\begin{align}
\hat{g}|_{z_1}=&\,h^{1/2}\,,\quad \hat{g}|_{-z_1}=h^{-1/2}\,,\quad
g_{(+)}|_{z_1}=h^{-1/2}\,,\quad g_{(-)}|_{-z_1}=h^{1/2}\,.
\end{align}
This gauge can be achieved by performing the 2d gauge transformation (\ref{2d-gauge}) with $\ch=h^{-1/2}$\,.
By using the notation (\ref{parameter-SG}), the explicit form is $h^{1/2}=\exp(\beta\phi \,T^3/2)$\,. 
The Lax form for this configuration is obtained as
\begin{align}\begin{split}
\cL_{\pm}=&\, h^{\mp 1/2}\pa_{\pm}h^{\pm 1/2} + im\frac{z\mp z_1}{z\pm z_1}h^{\pm 1/2}\Lambda_{\pm}h^{\mp 1/2}\,,\\
=&\,
\frac{im}{\sqrt{2}}\frac{z\mp z_1}{z\pm z_1}e^{\mp \frac{i}{2}\beta\phi}T^{+}
- \frac{im}{\sqrt{2}}\frac{z\pm z_1}{z\mp z_1}e^{\pm \frac{i}{2}\beta\phi}T^{-}
\pm \frac{\beta\pa_{\pm}\phi}{2}T^{3}\,,
\end{split}\end{align}
where we introduced $\Lambda_{\pm}=T^2$ and 
$
T^{\pm}:= \left( T^{1}\pm iT^{2} \right)/\sqrt{2}
$\,.
This is another Lax pair for the sine-Gordon model, and indeed the expression (\ref{2d-formula}) leads to the 2d action (\ref{SG-action}) of the sine-Gordon model.

\section{Conclusion and Discussion}\label{sec:conclusion}

We have presented a 4d CS theory with two order defects which lead to NATFTs.
In contrast to the case of the Faddeev-Reshetikhin model\cite{Fukushima:2020tqv}, the kinetic term of the 2d fields $\cG_{(\pm)}$ takes a quadratic form.
Although this form leads to the divergence at the classical level, an appropriate regularization (\ref{delta-smear}) and boundary conditions (\ref{gauge-boundaryA}) can resolve this divergence.
The $\mathbb{Z}_2$-grading structure for NATFTs is realized by imposing the constraint (\ref{lambda-constraint}) on the gauge field.

\medskip

A nontrivial potential term is possible only if the coefficients of the defect actions take the particular value $C_{(\pm)}=1/{z_1}$\,.
Otherwise, 
the equations of motion (\ref{V-EOM-2}) indicates that $m$ vanishes, and the theory just reduces to the WZW model.
The parameter $z_1\in\mathbb{R}$ characterizes integrable deformations. 
In the $z_1\to0$ limit, the simple poles degenerate into one double pole at $z=0$\,.
The condition (\ref{gauge-boundaryA}) then becomes just the Dirichlet boundary condition $A_{\pm}=0$\,, and the resulting 2d theory becomes a principal chiral model.

\medskip

A further extension of the derivation of the NATFTs is to lift up all of the flat directions.
This would be possible by considering a doubled 4d CS theory\cite{Stedman:2021wrw} (with disorder defects).
In this formalism, there is a pair of gauge fields $A$ and $B$\,, which take value in a Lie algebra $\mathfrak{g}$ and its subalgebra $\mathfrak{h}$\,, respectively.
The interaction term between $A$ and $B$ leads to ``gauged'' boundary conditions.
It is interesting to explore such a scenario in the presence of order defects.

%

\medskip

It is well-known that the B\"acklund transformation in 2d integrable field theories is realized as a spectral parameter-dependent gauge transformation which preserves the pole structure.
It would be interesting to discuss it in the context of a 4d CS theory.
After realizing it in terms of the 4d CS theory, one may get a hint on a new solution-generating technique in 2d integrable field theories.

\subsection*{Acknowledgments}

The work of O.\,F.\ was supported by Grant-in-Aid for JSPS Fellows No.~21J22806.
The work of J.S.\ was supported in part by Ministry of Science and Technology (project no.~110-2811-M-002-563), 
National Taiwan University.
The works of K.Y.\ was supported by 
JSPS Grant-in-Aid for Scientific Research (B) No.\,18H01214. This work was also supported in part by the JSPS Japan-Russia Research Cooperative Program.

\appendix

\section*{Appendix}

\section{Gauge choice (\ref{gauge-choice})}\label{app:gauge-choice}
In this appendix we confirm that the gauge choice (\ref{gauge-choice}) can be realized by performing the gauge transformation (\ref{gauge-trans}) and the 2d gauge transformation (\ref{2d-gauge}).

\medskip

The system is originally described in terms of $A$ and $\cG_{(\pm)}$\,. Then 
by performing the gauge transformation (\ref{formal-Lax}) characterized by $\hat{g}$ such that $A_{\bar{z}} = - (\partial_{\bar{z}}\hat{g})\hat{g}^{-1}$\,, the system is described 
by $\cL$ and $g_{(\pm)}$ with $\mathcal{L}_{\bar{z}} =0$\,.  

\medskip 

The gauge transformation (\ref{gauge-trans}) for the above setup leads to the relations: 
\begin{align}\begin{split}
A\mapsto A^{u}=&\,u\hat{g}\cL\hat{g}^{-1}u^{-1} - ud\hat{g}\cdot\hat{g}^{-1}u^{-1} - duu^{-1}\\
=&\, (u\hat{g})\cL (u\hat{g})^{-1} -d(u\hat{g})\cdot(u\hat{g})^{-1}\,,\\
\cG_{(\pm)}\mapsto \cG_{(\pm)}^{u}=&\, u\hat{g}g_{(\pm)}=(u\hat{g})g_{(\pm)}\,.
\end{split}
\label{gauge-hat}\end{align}
This expression means that $\hat{g}$ is replaced by $\hat{g}^u \equiv u \hat{g}$ in the transformation (\ref{formal-Lax})\,. 
Hence $\mathcal{L}$ and $g_{(\pm)}$ are invariant: 
\begin{align}
\cL\mapsto\cL^{u}=\cL\,,\qquad g_{(\pm)}\mapsto g_{(\pm)}^{u}=g_{(\pm)}\,.
\end{align}  
The gauge-fixing condition is not changed as well: 
\begin{eqnarray} 
\mathcal{L}_{\bar{z}}^u = \mathcal{L}_{\bar{z}} = 0\,. 
\end{eqnarray}

In the above discussion, we would like to take $u$ as
\begin{align}
u|_{\pm z_1}=(\hat{g}g_{(\pm)})^{-1}|_{\pm z_1} ~\left(=\cG_{(\pm)}^{-1}\big|_{\pm z_1}\right) \qquad
\Longleftrightarrow\quad
\cG_{(\pm)}^{u}\big|_{\pm z_1}=1\,. \label{cG=1}
\end{align}
As discussed in section \ref{sec:gauge-inv}, the gauge function $u$ needs to satisfy the boundary condition (\ref{u-derivative}) 
and the reality condition (\ref{u-parameter}). 
Hence this operation does not seems trivial at a glance. But this is indeed the case because 
$\cG_{(\pm)}$ should also satisfy the boundary condition (\ref{gauge-boundary}) and 
the grading constraint (\ref{lambda-constraint})\,. 
As for the boundary condition,  we see that 
\begin{align}
\pa_{\pm}u|_{\pm z_1}=0 \quad \mbox{and} \quad \left( \pa_{\pm}\cG_{(\pm)}^{-1}\right)\big|_{\pm z_1}=0\,.
\end{align}
Hence there is no obstruction for the choice (\ref{cG=1}). As for the reality condition, the constraints (\ref{reality-condition}) and (\ref{lambda-constraint}) indicate that $\cG_{(\pm)}$ must take values in $G_{0}$ because $z=\pm z_1$ are fixed points of $\mu_{\rm t}$ and $\lambda$\,.

\medskip

Note that the condition (\ref{cG=1}) is equivalent to $g_{(\pm)}^{u}|_{\pm z_1}=(\hat{g}^{u})^{-1}|_{\pm z_1}$\,. 
By employing the 2d gauge transformation (\ref{2d-gauge})\,, we can take 
\begin{eqnarray}
g_{(-)}^{u}|_{- z_1}=(\hat{g}^{u})^{-1}|_{- z_1}=1\,. 
\end{eqnarray}
Setting the remaining degrees of freedom as $\hat{g}^{u}|_{+z_1}=: h$\,, 
we have realized the following condtion: 
\begin{align}
\hat{g}^u|_{z_1}=h\,,\quad \hat{g}^u|_{-z_1}=1\,,\quad g_{(+)}^u|_{z_1}=h^{-1}\,,\quad g_{(-)}^u|_{-z_1}=1\,. 
\end{align}
After removing the superscript $u$\,, the expression (\ref{gauge-choice}) is obtained.

\end{document}